# Neural network interpretation using descrambler groups

Jake L. Amey, Jake Keeley, Tajwar Choudhury, and Ilya Kuprov*
University of Southampton, UK

## Abstract

The lack of interpretability and trust is a much-criticised feature of deep neural networks. In fully connected nets, the signalling between inner layers is scrambled because backpropagation training does not require perceptrons to be arranged in any particular order. The result is a black box; this problem is particularly severe in scientific computing and digital signal processing (DSP), where neutral nets perform abstract mathematical transformations that do not reduce to features or concepts.

We present here a group-theoretical procedure that attempts to bring inner layer signalling into a human-readable form, the assumption being that this form exists and has identifiable and quantifiable features – for example, smoothness or locality. We applied the proposed method to DEERNet (a DSP network used in electron spin resonance) and managed to descramble it. We found considerable internal sophistication: the network spontaneously invents a bandpass filter, a notch filter, a frequency axis rectifier, frequency division multiplexing, group embedding, spectral filtering regularisation, and a Fourier-like transform that maps harmonic functions into Chebyshev polynomials – in ten minutes of unattended training from a random initial guess.

## Introduction

Popular as the practice may be, simply training a neural net to perform a specific task is increasingly frowned upon – training is often just regression using the chain rule (*1*), and the resulting black box does not fit comfortably into the methodological framework (*2, 3*) of science and engineering. The concerns about deep neural nets are *interpretability* and *trust*, for which, at the moment, not even the definitions are settled. We can approximately define interpretability as *"the possibility of finding out why and how it works"* in the reductionist (*2*) and critical rationalist (*3*) sense, and trust as *"rigorous quantification of uncertainties in the output"*. Other related notions – intelligibility (*4*), algorithmic transparency (*5*), decomposability (*4*), attributability (*6*), transferability (*7*), and robustness (*8*) – may be viewed as aspects of these two general themes. Ultimately, the right answer for the right reasons is needed, accompanied by a measure of certainty (*9*).

A fully connected feed-forward artificial neural network with an input vector $\mathbf{x}$ and an output vector $\mathbf{y}$ is equivalent to the following function:

$$\mathbf{y} = F_n \mathbf{W}_n F_{n-1} \mathbf{W}_{n-1} \cdots F_1 \mathbf{W}_1 \mathbf{x} \tag{1}$$

where $\mathbf{W}_k$ are weight matrices, $F_k$ are non-linear activation functions, and bias vectors are not specified because in this case they are equivalent to having one extra input line. It is convenient to supply and receive arrays of input and output vectors; those will be denoted $\mathbf{X}$ and $\mathbf{Y}$ respectively. The horizontal dimension of $\mathbf{W}_1$ is ordered in the same way as columns of $\mathbf{X}$, the vertical dimension of $\mathbf{W}_n$ is ordered in the same way as columns of $\mathbf{Y}$, all other dimensions of $\mathbf{W}_k$ are not ordered because backpropagation training does not require them to be, and the initial guess is random. We call such weight matrices "scrambled": they are two linear transformations – one from the left and one from the right – away from a representation with ordered input and output.

## Descrambler group

We assume that neural networks are interpretable – that, for each layer *k*, a transformation $\mathbf{P}$ exists that brings the signal array $F_k \mathbf{W}_k \cdots F_1 \mathbf{W}_1 \mathbf{X}$ into a form that clarifies, to a competent human, the function if the preceding layers. We call this a "descrambling" transformation. The activation functions are not varied in the training process, and therefore this transformation must be applied to the weight matrices and judged on the output signals, for which some interpretability metric must be designed.

The transformation should be linear, so that linear combinations of signals are descrambled consistently. Information should not be lost, and therefore the transformation must be invertible. There also exists a unit transformation that does nothing. That is the definition of a group which we will call the descrambler group. At the *k*[th] layer of the network, it must be a subgroup of the general linear group of all automorphisms of a $d_k$-dimensional vector space, where $d_k$ is the output dimension of the layer. It must also be a supergroup of the permutation group of $d_k$ perceptrons within the layer. Of those, $SO(d_k)$ – the connected group of all proper orthogonal transformations of a $d_k$-dimensional vector space – is particularly promising, because physical signals are often defined up to an orthogonal transformation (*e.g.* cosine transform), and because the elements of $SO(d_k)$ are continuous and differentiable functions of a finite number of real parameters.

The act of wiretapping the network at a particular layer then consists of inserting a unit operator $\mathbf{P}^{-1}\mathbf{P}$ before or after the weight matrix:

$$\mathbf{Y} = F_n \mathbf{W}_n \cdots F_k \mathbf{P}^{-1} \mathbf{P} \mathbf{W}_k \cdots F_1 \mathbf{W}_1 \mathbf{X}$$
$$or \qquad (2)$$
$$\mathbf{Y} = F_n \mathbf{W}_n \cdots \mathbf{P}^{-1} \mathbf{P} F_k \mathbf{W}_k \cdots F_1 \mathbf{W}_1 \mathbf{X}$$

and maximising or minimising such a function $\Lambda$ of $\mathbf{P} \mathbf{W}_k \cdots F_1 \mathbf{W}_1 \mathbf{X}$ or $\mathbf{P} F_k \mathbf{W}_k \cdots F_1 \mathbf{W}_1 \mathbf{X}$ as would quantify the features forming the basis of the interpretation, *e.g.*

$$\mathbf{P} = \arg \begin{Bmatrix} \min \\ \max \end{Bmatrix} \Lambda \left( \mathbf{P} \mathbf{W}_k \cdots F_1 \mathbf{W}_1 \mathbf{X} \right) \qquad (3)$$

Much creativity may be needed to construct that function: it must take a signal and return a quantitative figure of merit for some problem- or domain-specific definition of "interpretable". This could be a function involving measures of smoothness, periodicity, monotonicity, locality, autocorrelation, Shannon entropy, deviation from the expected statistics on luminosity and chromaticity, *etc.*

In our context – digital signal processing – the smoothness of the time domain signal is a good metric: a transformation that makes every intermediate signal across a large input library simultaneously smooth is also likely to make them physically meaningful. We have chosen Tikhonov smoothness – the squared Euclidean norm of the second derivative – as the metric to be minimised:

$$\Lambda(\mathbf{v}) = \|\mathbf{D}\mathbf{v}\|^2 \tag{4}$$

where $\mathbf{D}$ is a representation of the second derivative operator on a finite grid with $d_k$ points; we use Fourier spectral differentiation matrices (*10*).

When multiple output vectors are concatenated into a matrix, the sum of squares of their Euclidean norms is the square of the Frobenius norm of that matrix. Therefore, applied to the output array of the $k^{\text{th}}$ layer of the network, Tikhonov smoothness criterion becomes:

$$\mathbf{P} = \arg\min_{\mathbf{P}} \|\mathbf{D}\mathbf{P}\mathbf{W}_k \cdots F_1 \mathbf{W}_1 \mathbf{X}\|_F^2, \qquad \|\mathbf{A}\|_F^2 = \text{Tr}(\mathbf{A}^T \mathbf{A}) \tag{5}$$

where $\|\_\|_F$ denotes Frobenius norm, and $\mathbf{X}$ is a large enough array of input vectors (in practice, the entire training database). Importantly, Equation (5) is not equivalent to smoothing the columns of the weight matrix by minimising $\|\mathbf{D}\mathbf{P}\mathbf{W}_k\|_F^2$. This is because only smoothness in the outgoing data is sought – a weaker requirement. Equation (5) is also a weaker requirement than placing a Tikhonov penalty on the weight matrix at the training stage – an interpretable matrix need not itself be smooth, it only needs to produce intelligible signalling. Accordingly, the metrics being optimised in Equations (3) and (5) refer not to the weight matrices, but to the intermediate signal arrays.

In the absence of constraints, the obvious solution to Equation (5) is $\mathbf{P} = 0$ – this is why a group-theoretical approach is needed where $\mathbf{P}$ is generated by the Lie algebra of the descrambler group, and thus constrained to be non-singular. However, the usual exponential map $\mathbf{P} = \exp(\mathbf{Q})$ has expensive derivatives and numerical accuracy problems in finite precision arithmetic. We have therefore opted for a different connection between $SO(d_k)$ and its algebra, called Cayley transform (*11*):

$$\mathbf{P} = \frac{1 - \mathbf{Q}}{1 + \mathbf{Q}} \tag{6}$$

where the numerator acts first, and $\mathbf{Q}$ is an antisymmetric matrix. Cayley transform is less sensitive to extreme eigenvalues than the matrix exponential. It is also easier to differentiate (see Supplementary Information) with respect to $\mathbf{Q}$. The general case remains as in Equation (3), *e.g.*

$$\mathbf{Q} = \arg\left\{\begin{array}{c}\min\\\max\end{array}\right\} \Lambda\left(\frac{1 - \mathbf{Q}}{1 + \mathbf{Q}} \mathbf{W}_k \cdots F_1 \mathbf{W}_1 \mathbf{X}\right) \tag{7}$$

and the specific case of hoping for Tikhonov smoothness in the output of the weight matrix of a particular layer is equivalent to minimising

$$\eta_T(\mathbf{Q}) = \left\|\mathbf{D}\frac{1 - \mathbf{Q}}{1 + \mathbf{Q}} \mathbf{W}_k \cdots F_1 \mathbf{W}_1 \mathbf{X}\right\|_F^2 \tag{8}$$

with respect to the real antisymmetric matrix $\mathbf{Q}$. The gradient $\partial \eta_T / \partial \mathbf{Q}$ is cheap (Section S1 of the Supplementary Information), meaning that quasi-Newton optimisers like L-BFGS (*12*) may be used.

Memory utilisation is likewise not a problem – Frobenius norm-square is additive with respect to the columns of $\mathbf{X}$, which may be fed into the calculation one by one, or in batches. Thus, if a network can be trained on some hardware, it can also be descrambled on the same hardware.

## Fredholm solver networks and DEERNet

Consider the trajectory $\gamma(x,t)$ for a property $\gamma$ in a quantum system with a parameter $x$. When the sample contains an ensemble of systems with a probability density $p(x)$ in that parameter, the result $\Gamma(t)$ of the ensemble average measurement is given by Fredholm's integral (*13*):

$$\Gamma(t) = \int p(x)\gamma(x,t)dx \qquad (9)$$

where $\gamma(t,x)$ is sometimes called the "kernel"; its exact form depends on the physics of the problem. This integral is at the heart of applied quantum mechanics, used (directly or indirectly) for interpretation of any physical experiment by a model with distributed parameters. Given an experimentally measured $\Gamma(t)$, extracting $p(x)$ is hard: without regularisation, this is an ill-posed problem (*14*), and regularisation brings in a host of other complications (*15*). Deep neural networks perform unexpectedly well here (*16*), but no explanation exists as to why.

Our instance of this problem came from structural biology: molecular distance determination using DEER – double electron-electron resonance (*17*). We generated a large database of realistic distance distributions and complications (noise, baseline, *etc.*) and converted them into what the corresponding experimental data would look like. Acting out of curiosity, we put together a fully connected feedforward neural net and trained it to perform the inverse transformation – from noisy and distorted $\Gamma(t)$ back into $p(x)$. Because the problem is ill-posed, this was not supposed to be possible. The network did it anyway (Figure 1), and matched the best regularisation solver there is (*16*).

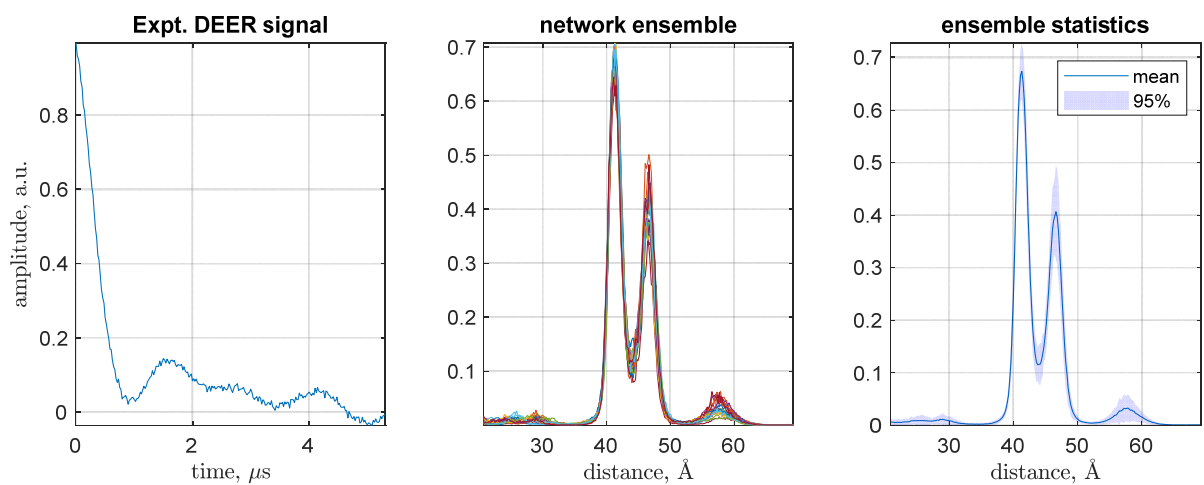

*Figure 1.* A typical double electron-electron resonance (DEER) dataset from structural biology work. The left panel shows the electron spin echo modulation between two iodocateamido-PROXYL spin labels attached to the incoming cysteines in the V96C, I143C double mutant of Light Harvesting Complex II in n-octyl-β-D-glucoside micelles (*18*). The middle panel shows distance probability densities returned by an ensemble of independently trained neural networks in DEERNet (*16*). The right panel contains statistics across the neural network ensemble.

Mathematicians had looked at such things – neural network "surrogate" solutions to Fredholm equations had been attempted (*19*), and accuracy bounds are available (*20*). In 2013, Jafarian and Nia proposed a feed-back network built around a Taylor expansion of the solution (*21*); a feed-forward network proposition was published in 2012 by Effati and Buzhabadi (*22*). Both groups reported accurate solutions (*21, 22*), but neither looked at applications, or asked the question about how a neural network actually manages to regularise the problem.

Given the precarious interpretability of quantum mechanics itself, demanding it from a neural network trained on quantum mechanics may seem unreasonable. However, this case is an exception: electron spin dynamics is very well understood, and the networks in question are uncommonly small – only 256 perceptrons wide, with at most ten layers (*16*). We have therefore picked DEERNet as a test case for the descrambler group method. The simple and clear case involving two fully connected layers is discussed here, the case with three fully connected layers is in the Supplementary Information.

### Descrambling DEERNet

The simplest DEERNet has the following layer structure: vector input > fully connected > sigmoidal function > fully connected > logsig function > vector regression. The logsig activation function is necessary to ensure that the output (which has a physical meaning of probability density) stays positive. The input and the output are 256 elements wide, but the link dimension may be reduced to 80 by eliminating insignificant singular values (*16*) from the weight matrices of fully connected layers. The input dimension of $\mathbf{W}_1$ is time-ordered (Figure 1, left panel), the output dimension of $\mathbf{W}_2$ is distance-ordered (Figure 1, right panel), but the link dimension connecting $\mathbf{W}_1$ and $\mathbf{W}_2$ is scrambled.

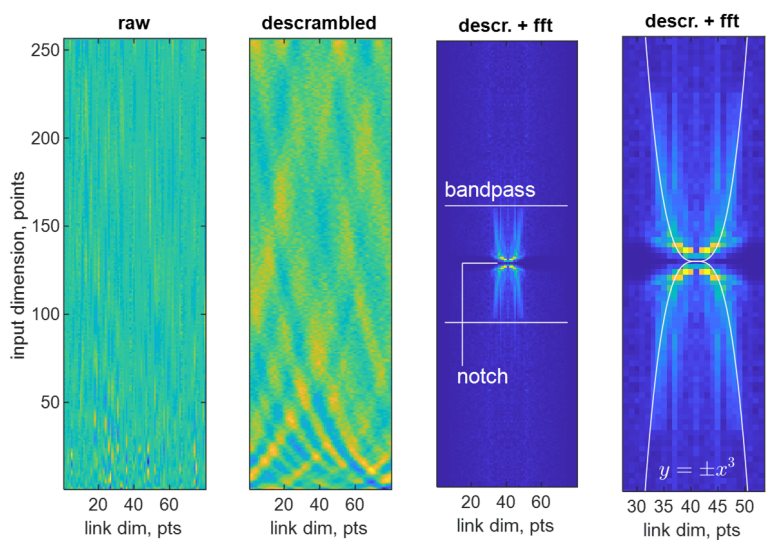

*Figure 2. Spontaneous emergence of a sophisticated digital filter in the first fully connected layer of a DEERNet (16) neural network. From left to right: raw weight matrix of the input layer, descrambled weight matrix, symmetrised absolute value 2D Fourier transform of the descrambled weight matrix, and a zoom into the central portion of that Fourier transform with a cubic curve overlaid. The layer applies a low pass filter to remove high-frequency noise seen in the left panel of Figure 1, a notch filter at zero frequency to remove the non-oscillatory baseline, and also performs frequency axis rectification from cubic to linear – apparently, to account for the fact that the quantum beat frequency in DEER (17) is an inverse cubic function of the distance between the spins.*

Applying the descrambler group method to minimise the second derivative norm of the output of $\mathbf{W}_1$ (Figure 2, left) reveals a rich structure (Figure 2, middle left) – the interlocking wave pattern indicates that some kind of frequency conversion is being performed. Inserting forward ($\mathbf{F}_+$) and backward ($\mathbf{F}_-$) Fourier transforms into the corresponding equation:

$$\mathbf{y} = \mathbf{W}\mathbf{x} \quad \Rightarrow \quad \mathbf{F}_+\mathbf{y} = \mathbf{F}_+\mathbf{W}\mathbf{F}_-\mathbf{F}_+\mathbf{x} \tag{10}$$

demonstrates that the input signal frequency spectrum $\mathbf{F}_+\mathbf{x}$ is connected to the output signal frequency spectrum $\mathbf{F}_+\mathbf{y}$ by $\mathbf{F}_+\mathbf{W}\mathbf{F}_-$ matrix. Computing and plotting this matrix (Figure 2, middle right) reveals the function of the first fully connected layer: it applies a low-pass filter to eliminate high-frequency noise, a notch filter at the zero frequency to eliminate the non-oscillatory baseline, and performs frequency axis rectification from cubic to linear within the bandwidth of the low-pass filter (Figure 2, right). The latter operation appears to reflect the fact that the quantum beat frequency in kernel function of DEER depends on the cube of the distance (*17*):

$$\gamma(r,t) = \sqrt{\frac{\pi}{6Dt}}\left[\cos[Dt]\mathrm{FrC}\left[\sqrt{\frac{6Dt}{\pi}}\right] + \sin[Dt]\mathrm{FrS}\left[\sqrt{\frac{6Dt}{\pi}}\right]\right]$$

$$D = \frac{\mu_0}{4\pi}\frac{\gamma_1\gamma_2\hbar}{r^3}; \qquad \mathrm{FrC}(x) = \int_0^x \cos(t^2)dt \qquad \mathrm{FrS}(x) = \int_0^x \sin(t^2)dt \tag{11}$$

where $\gamma_{1,2}$ are magnetogyric ratios of the two electrons and $r$ is the inter-electron distance. All three operations are linear filters; the network managed to pack them into one layer. The function of the layer is now clear – baseline elimination, noise elimination, and signal preprocessing.

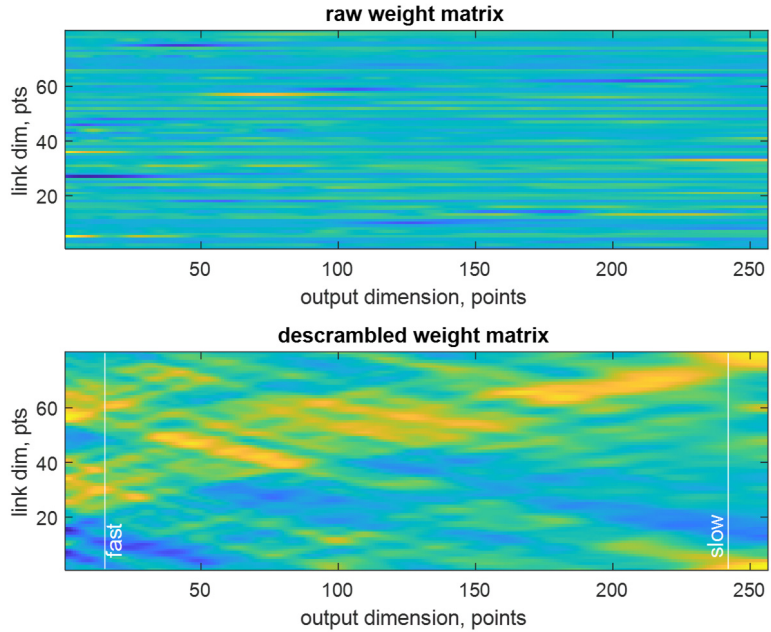

*Figure 3.* Spontaneous emergence of a Fourier-like time-frequency transform in the second fully connected layer of a DEERNet (16) neural net. Once the descrambler group method is applied to the raw weight matrix (top panel), its vertical dimension becomes interpretable (bottom panel). Although the matrix is not exactly a Fourier transform, it does appear to map faster oscillations into shorter distances and slower oscillations into longer distances – the horizontal dimension of the output layer weight matrix is the distance axis of the output.

Since the preceding layer is a digital filter that keeps the problem in the time domain, some form of time-to-frequency domain transformation is expected in the weight matrix of the second fully connected layer (Figure 3, top). Applying the descrambler group method to minimise simultaneously the second derivative norm of the output of the transfer function of the previous layer, and the second derivative along the link dimension of $\mathbf{W}_2$, reveals a Fourier-like transform (Figure 3, bottom) where fast oscillations are mapped into short distances and slow oscillations into long distances.

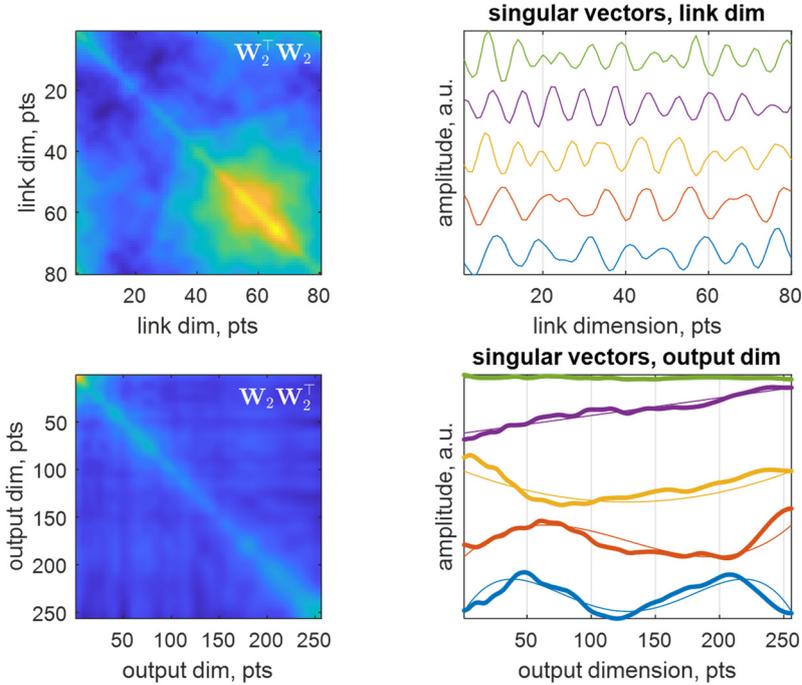

*Figure 4.* Spontaneous emergence of frequency division multiplexing and Chebyshev polynomials in the second fully connected layer of a DEERNet (16) neural net. Descrambling the link dimension reveals an approximately orthogonal (top left panel) conjugate signal library that singular value decomposition shows to be distorted sinusoids (top right panel). The output signal library also appears to be approximately orthogonal (bottom left panel); singular value decomposition reveals spontaneous emergence of distorted Chebyshev polynomials as the entries of that library (bottom right panel).

A more detailed inspection reveals that both the rows and the columns of the descrambled $\mathbf{W}_2$ are approximately orthogonal (Figure 4, top left and bottom left). This prompted us to run singular value decomposition (SVD) to find out what the descrambled weight matrix expects to receive and to send out. SVD is useful after descrambling because its structure

$$\mathbf{W} = \mathbf{U}\mathbf{S}\mathbf{V}^\dagger \quad (12)$$

naturally breaks the weight matrix down into an orthogonal set of conjugate signals that it expects to receive (columns of $\mathbf{V}$), amplification coefficients for the signals received (elements of the diagonal matrix $\mathbf{S}$), and an orthogonal set of signals that it expects to send out (columns of $\mathbf{U}$) with those amplification coefficients in response to each of the signals it has received. SVD revealed that the conjugate input signals are sinusoids, slightly distorted, likely due to imperfect training (Figure 4, top right) – the network apparently invented frequency division multiplexing. The output signals appear to be distorted Chebyshev polynomials (Figure 4, bottom right).

Exactly why the network went specifically for Chebyshev polynomials is unclear, but they provide the explanation of how regularisation is done inside DEERNet: the ranks of the Chebyshev polynomials seen in the output signal library are smaller than the ranks that can in principle be digitised on the 256-point output grid. Thus, a degree of smoothness is enforced in the output signal – the procedure is reminiscent of spectral filtering regularisation. It also has a modicum of elegance: the log-sigmoidal transfer function of the output layer in DEERNet neatly converts Chebyshev polynomials into patterns of peaks, as required by the physics of the problem (*17*).

The network is now completely interpreted: the first fully connected layer is a digital filter that performs denoising, baseline elimination, and frequency axis rectification, and then sends the signal, in a frequency-multiplexed form, to the second fully connected layer, which performs a regularised time-to-frequency transformation into Chebyshev polynomials that the final log-sigmoidal transfer function converts into the patterns of peaks seen in the right panel of Figure 1.

Further examples (a DEERNet with three fully connected layers and a network designed to eliminate additive noise from human voice recordings) may be found in the Supplementary Information.

## Conclusions and outlook

The descrambler group method made it possible to interpret the functioning of a fully connected neural network. During its training, a simple DEERNet appears to have invented: a bandpass filter, a notch filter, a frequency axis rectifier, frequency division multiplexing, spectral filtering regularisation, and a Fourier-like transform that maps harmonic functions into Chebyshev polynomials. As far as we can tell, a deeper DEERNet (Supplementary Information) also invented group embedding.

That these tiny networks should develop this amount of instantly recognisable mathematics and communications engineering in ten minutes of unattended training from a random initial guess is unexpected. The functionality appears to be localised and readable to humans, meaning that reductionism (*2*) and critical rationalism (*3*) need not be abandoned, at least for the smaller neural networks. An ironic observation is that the act of interpreting the inner working of a static neural net apparently obviates the need for it: the same filters and transforms may now be applied rationally – a functional replica of the two-layer DEERNet may be found in the Supplementary Information.

A key strength of the descrambler group method is its applicability to fully connected layers – those are harder to interpret than convolutional layers, which inherit partial order from the convolution stride. Due to their importance in image processing, the existing interpretability scoring methods tend to focus on convolutional nets (*23*). Other established methods, for example concept activation vectors (*24*) and saliency maps (*25*), are specific to object detection and classification networks where identifiable concepts exist. This is not necessarily the case in digital signal processing networks like DEERNet that apply abstract non-linear maps between vector spaces.

There is also a difference between finding out what a network does, and finding out how. The approach presented here is more firmly grounded in formal mathematics than many of the current Explainable AI techniques, of which some are variations of "poke it with a stick and see what happens"

empiricism – that is what differentiating the network with respect to an input, output, or a parameter fundamentally is: increment something, look at the change in behaviour. Much of the prior art deals specifically with expert, recommender, and classifier systems, and thus with extracting rule lists, decision trees, and taxonomies (*26*). None of that is relevant to networks that evolve abstract mathematical transforms between signal spaces – descrambler groups offer an opportunity here, because they run on generic mathematical properties of those signals.

The definition of the descrambling target functional in general is entirely at the user's discretion – the functional in Equation (8) is only one of many possibilities. For example, in situations when frequency domain data is expected at both the input and the output of an acoustic filter network, it is reasonable to seek a transformation of the output space that makes output signals maximally similar to the input ones. In that representation, the weight matrix $\mathbf{W}$ is expected to be diagonally dominant; this may be achieved by seeking an orthogonal transformation of the output space that achieves maximum diagonal sum (MDS) or maximum diagonal norm square (MDNS) for the weight matrix:

$$\eta_{\mathrm{MDS}}(\mathbf{Q}) = \mathrm{Tr}[\mathbf{PW}], \qquad \eta_{\mathrm{MDNS}}(\mathbf{Q}) = \left\|\mathrm{diag}(\mathbf{PW})\right\|_2^2, \qquad \mathbf{P} = \frac{\mathbf{1}-\mathbf{Q}}{\mathbf{1}+\mathbf{Q}} \qquad (13)$$

An example of using this approach for a network designed to remove additive noise from human speech is given in the Supplementary Information.

A discomfiting aspect of the present work is the amount of domain-specific expertise that was required to recognise the functionality of the descrambled weight matrices. It could be argued that the matrix in the second panel of Figure 2 is still uninterpretable to a non-specialist. That is an improvement though – the matrix in the first panel was uninterpretable to everyone. We recommend a staged approach: the role of each layer should first be established empirically using the prior art cited above, and then the weight matrix descrambled to find out the implementation details. Still, the availability of such methods opens a way to deeper study of neural networks because the training stage can now be followed by the interpretation stage. So far, we have only seen our networks invent the mathematics that is known to humans. It is possible that, at some point, previously unknown mathematics would make an appearance: neural nets can likely be mined for new knowledge.

### Acknowledgements
We thank Steven Worswick, Gunnar Jeschke, Daniella Goldfarb, Thomas Prisner, and Robert Bittl for valuable discussions, as well as MathWorks for the excellent technical support. **Funding:** the authors acknowledge the support from EPSRC through an Impact Acceleration Grant, from Leverhulme Trust (RPG-2019-048), from MathWorks through a PhD studentship donation, and also the use of the IRIDIS High Performance Computing Facility and associated support services at the University of Southampton. **Author contributions:** IK came up with the idea of the descrambler group and performed the analytical mathematics; JLA derived the gradient expression and implemented the descrambling algorithm in Matlab; JK descrambled the middle weight matrix of the DEERNet with three fully connected layers; TC trained and descrambled the denoising network; the analysis of descrambled weight matrices and the writing of the manuscript were done jointly. **Competing interests:** the authors declare that they have no competing interests. **Data and materials availability:** the source code of DEERNet is available as a part of the open-source Spinach package (http://spindynamics.org).

# Neural network interpretation using descrambler groups

Jake L. Amey, Jake Keeley, Tajwar Choudhury, and Ilya Kuprov*
University of Southampton, UK

**Electronic Supplementary Information**

## S1. Gradient of the Tikhonov descrambling functional

The gradient of the functional in Equation (8) of the main text:

$$\eta_\text{T}(\mathbf{Q}) = \left\| \mathbf{D} \frac{\mathbf{1} - \mathbf{Q}}{\mathbf{1} + \mathbf{Q}} \mathbf{W}_k \cdots F_1 \mathbf{W}_1 \mathbf{X} \right\|_\text{F}^2 \tag{S.1}$$

may be obtained using matrix differentiation rules. The weighted array of column vector signals arriving from the preceding layers will be abbreviated as:

$$\mathbf{S} = \mathbf{W}_k F_{k-1} \mathbf{W}_{k-1} \cdots F_1 \mathbf{W}_1 \mathbf{X} \tag{S.2}$$

With the result that the descrambling functional at layer $k$ becomes:

$$\eta_\text{T}(\mathbf{Q}) = \left\| \mathbf{DPS} \right\|_\text{F}^2 \qquad \mathbf{P} = \frac{\mathbf{1} - \mathbf{Q}}{\mathbf{1} + \mathbf{Q}} \tag{S.3}$$

where we chose $\mathbf{D}$ to be a second Fourier derivative (*1*) matrix. Using the chain rule:

$$\frac{\partial \eta_\text{T}}{\partial Q_{ij}} = \text{Tr}\left[ \left(\frac{\partial \eta_\text{T}}{\partial \mathbf{P}}\right)^\text{T} \frac{\partial \mathbf{P}}{\partial Q_{ij}} \right] = \sum_{kl} \left[\frac{\partial \eta_\text{T}}{\partial \mathbf{P}}\right]_{lk} \frac{\partial P_{lk}}{\partial Q_{ij}} \tag{S.4}$$

The derivative of $\mathbf{P}$ with respect to an element of $\mathbf{Q}$ is another instance of the chain rule:

$$\begin{aligned}
\frac{\partial P_{lk}}{\partial Q_{ij}} &= \left[ \frac{\partial}{\partial Q_{ij}} \left[ (\mathbf{I} + \mathbf{Q})^{-1}(\mathbf{I} - \mathbf{Q}) \right] \right]_{lk} = \\
&= \left[ \frac{\partial (\mathbf{I} + \mathbf{Q})^{-1}}{\partial Q_{ij}}(\mathbf{I} - \mathbf{Q}) + (\mathbf{I} + \mathbf{Q})^{-1} \frac{\partial (\mathbf{I} - \mathbf{Q})}{\partial Q_{ij}} \right]_{lk} = \\
&= \sum_m \frac{\partial \left[(\mathbf{I} + \mathbf{Q})^{-1}\right]_{lm}}{\partial Q_{ij}}[\mathbf{I} - \mathbf{Q}]_{mk} - \sum_n \left[(\mathbf{I} + \mathbf{Q})^{-1}\right]_{ln} \frac{\partial Q_{nk}}{\partial Q_{ij}}
\end{aligned} \tag{S.5}$$

The last derivative is $\partial Q_{nk}/\partial Q_{ij} = \delta_{ni}\delta_{kj}$, and the derivative of the inverse matrix is:

$$\frac{\partial \left[ (\mathbf{I}+\mathbf{Q})^{-1} \right]_{lm}}{\partial Q_{ij}} = -\left[ (\mathbf{I}+\mathbf{Q})^{-1} \right]_{li} \left[ (\mathbf{I}+\mathbf{Q})^{-1} \right]_{jm} \qquad (S.6)$$

This eliminates all derivatives and all explicit sums from the right-hand side:

$$\begin{aligned}\frac{\partial P_{lk}}{\partial Q_{ij}} &= -\sum_m \left[ (\mathbf{I}+\mathbf{Q})^{-1} \right]_{li} \left[ (\mathbf{I}+\mathbf{Q})^{-1} \right]_{jm} \left[ \mathbf{I}-\mathbf{Q} \right]_{mk} - \sum_n \left[ (\mathbf{I}+\mathbf{Q})^{-1} \right]_{ln} \delta_{ni} \delta_{kj} \\ &= -\left[ (\mathbf{I}+\mathbf{Q})^{-1} \right]_{li} \left[ (\mathbf{I}+\mathbf{Q})^{-1}(\mathbf{I}-\mathbf{Q}) \right]_{jk} - \left[ (\mathbf{I}+\mathbf{Q})^{-1} \right]_{li} \delta_{kj} \end{aligned} \qquad (S.7)$$

Using the definitions of $\mathbf{P}$ and $\mathbf{I}$ yields further simplifications:

$$\frac{\partial P_{lk}}{\partial Q_{ij}} = -\left[ (\mathbf{I}+\mathbf{Q})^{-1} \right]_{li} \left[ \mathbf{I}+\mathbf{P} \right]_{jk} \qquad (S.8)$$

Inserting this into Equation (S.4) produces:

$$\frac{\partial \eta_\mathrm{T}}{\partial Q_{ij}} = -\sum_{kl} \left[ (\mathbf{I}+\mathbf{Q})^{-1} \right]_{li} \left[ \frac{\partial \eta_\mathrm{T}}{\partial \mathbf{P}} \right]_{lk} \left[ \mathbf{I}+\mathbf{P} \right]_{jk} \qquad (S.9)$$

The explicit sum can now be collapsed:

$$\frac{\partial \eta_\mathrm{T}}{\partial Q_{ij}} = -\left[ (\mathbf{I}+\mathbf{Q})^{-\mathrm{T}} \frac{\partial \eta_\mathrm{T}}{\partial \mathbf{P}} (\mathbf{I}+\mathbf{P})^\mathrm{T} \right]_{ij} \qquad (S.10)$$

The derivative of $\eta$ with respect to $\mathbf{P}$ is obtained using the Frobenius norm differentiation rule:

$$\frac{\partial \eta_\mathrm{T}}{\partial \mathbf{P}} = 2\mathbf{D}^\mathrm{T} \mathbf{D} \mathbf{P} \mathbf{S} \mathbf{S}^\mathrm{T} \qquad (S.11)$$

The final result is:

$$\frac{\partial \eta_\mathrm{T}}{\partial \mathbf{Q}} = -2(\mathbf{I}+\mathbf{Q})^{-\mathrm{T}} \left[ \mathbf{D}^\mathrm{T} \mathbf{D} \right] \mathbf{P} \left[ \mathbf{S} \mathbf{S}^\mathrm{T} \right] (\mathbf{I}+\mathbf{P})^\mathrm{T} \qquad (S.12)$$

Numerical evaluation of both the functional and the gradient may be accelerated by pre-computing the terms enclosed in square brackets.

## S2. Gradients of maximum diagonality descrambling functionals

The gradients of the two functionals (maximum diagonal sum and maximum diagonal norm squared) in Equation (13) of the main text

$$\eta_\mathrm{MDS} = \mathrm{Tr}[\mathbf{PW}], \qquad \eta_\mathrm{MDNS} = \left\| \mathrm{diag}(\mathbf{PW}) \right\|_2^2, \qquad \mathbf{P} = \frac{\mathbf{1}-\mathbf{Q}}{\mathbf{1}+\mathbf{Q}} \qquad (S.13)$$

are obtained in a similar way to the above:

$$\frac{\partial \eta_{\text{MDS}}}{\partial \mathbf{Q}} = -(\mathbf{1}+\mathbf{Q})^{-T}\mathbf{W}^{T}(\mathbf{1}+\mathbf{P})^{T}$$
$$\frac{\partial \eta_{\text{MDNS}}}{\partial \mathbf{Q}} = -2(\mathbf{1}+\mathbf{Q})^{-T}\left[\mathbf{W}^{T}\odot\text{diag}(\mathbf{PW})\right](\mathbf{1}+\mathbf{P})^{T} \quad (S.14)$$

where $\odot$ denotes element-wise multiplication.

## S3. DEERNet topology and training

A training database containing $10^5$ DEER traces was generated as we previously described (*2*); networks of the following general topology were trained using scaled conjugate gradient backpropagation on an NVidia Titan V card using Matlab R2020a Machine Learning Toolbox (*3*); technical details may be found in the same paper (*2*). The input layer and the inner layers of DEERNet use tangent sigmoidal activation functions, and the output layer uses the log-sigmoidal activation function to ensure that the output cannot become negative. The dimension of the inner layer weight matrices is chosen based on the weight matrix rank analysis (*2*).

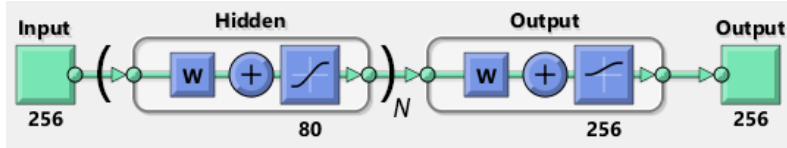

## S4. Replicating DEERNet using digital signal processing

To confirm the correctness of the DEERNet functionality interpretation presented in the main text, we have assembled a combination of digital filters that replicates the functionality of the first fully connected layer, and a time-distance transformation that replicates the functionality of the second one.

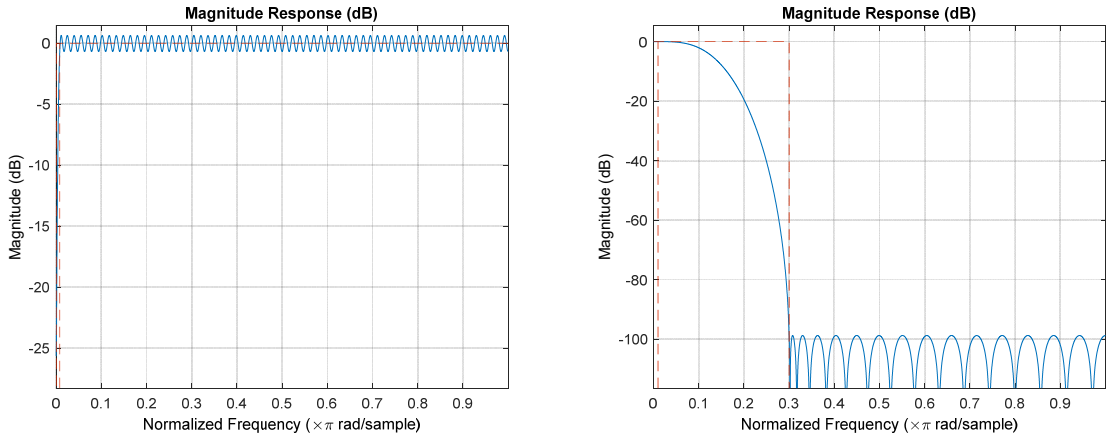

*Figure S1.* Digital filters used in the re-creation of the functionality of the first fully connected layer of DEERNet. **Left panel**: notch filter at zero frequency, implemented as order 256 direct-form FIR high-pass filter with passband edge at 0.008 and stopband edge at 0.001 normalised frequency units. **Right panel:** order 32 direct-form FIR low pass filter with passband edge at 0.01 and stopband edge at 0.3 normalised frequency units. Filters created and analysed using the Signal Processing Toolbox of Matlab R2020a.

To emulate the first fully connected layer, we used standard FIR filters with pass and reject bands (Figure S1) chosen to correspond approximately to the patterns seen in Figure 2 of the main text. The frequency rectification transform and the regularised time-distance transform we had found in the second fully connected layer are both linear, and were therefore combined into one matrix $\mathbf{T}$ that was obtained as a regularised pseudoinverse:

$$\mathbf{T}\begin{bmatrix}\mathbf{f}_1 & \cdots & \mathbf{f}_n\end{bmatrix} = \begin{bmatrix}\mathbf{p}_1 & \cdots & \mathbf{p}_n\end{bmatrix}, \qquad n \gg \dim(\mathbf{T})$$
$$\Downarrow \qquad\qquad\qquad\qquad\qquad\qquad\qquad (S.15)$$
$$\|\mathbf{TF}-\mathbf{P}\|_F^2 + \lambda\|\mathbf{T}\|_F^2 = \min \quad \Rightarrow \quad \mathbf{T} = \left[\left(\mathbf{FF}^T + \lambda\mathbf{1}\right)^{-1}\left(\mathbf{FP}^T\right)\right]^T$$

where $\mathbf{p}_k$ are linearly independent distance probability density distributions presented as vectors on a finite grid, and $\mathbf{f}_n$ are the corresponding solutions of Eq (9) of the main text, also discretised on a finite grid. The regularisation parameter $\lambda$ was obtained using the L-curve method (*4*). Although some parameters (filter orders and bands, pseudoinverse regularisation factor) were chosen empirically, they all now have a clear rational interpretation – thus, a *physically meaningful* data processing method was obtained from a descrambler group interpretation of a neural network.

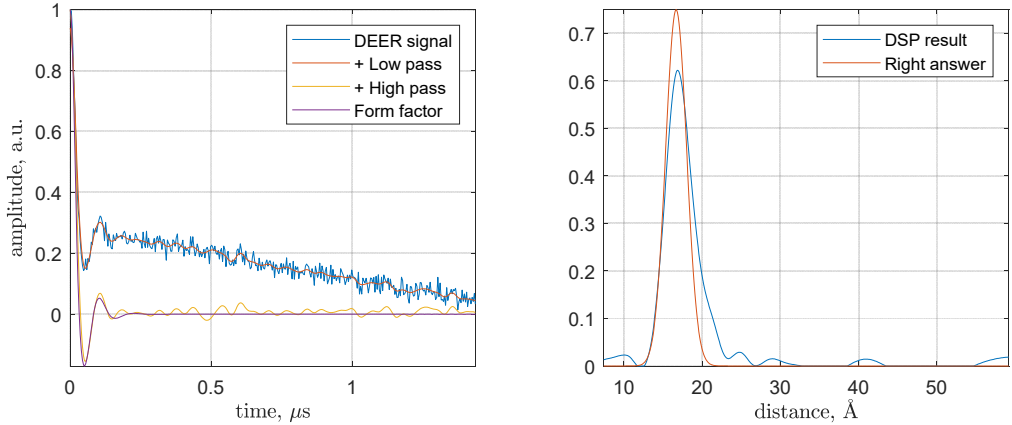

*Figure S2. An example DEER processing run using the rational digital signal processing replica of DEERNet. The calculation starts with a realistic randomly generated DEER dataset (left panel, blue trace) for which the correct answer is known. The low pass filter eliminates the noise (left panel, red trace), and the notch filter at zero eliminates the baseline (left panel yellow trace). Up to the noise, the result matches the known right answer at this stage (left panel, purple trace). The time-frequency transform yields a distance pattern in reasonable agreement with the known right answer (right panel).*

The performance of the rationally constructed transform sequence is illustrated in Figure S2 – it is clear that the performance is similar to the performance shown by the neural network ensemble in Figure 1 of the main text. Across a large database of inputs that we had inspected, the rational method does require occasional pass and reject band adjustments in the digital filters to match the performance of the network, but those adjustments always have a physical explanation.

## S5. The extra layer of a deeper DEERNet

Descrambling the DEERNet featuring three fully connected layers revealed that the first one applies a similar combination of digital filters to those shown in Figure 2 of the main text, and the last one is a time-distance transform similar to the one in Figures 3 and 4. However, the middle fully connected layer turned out to have unexpected mathematical depth.

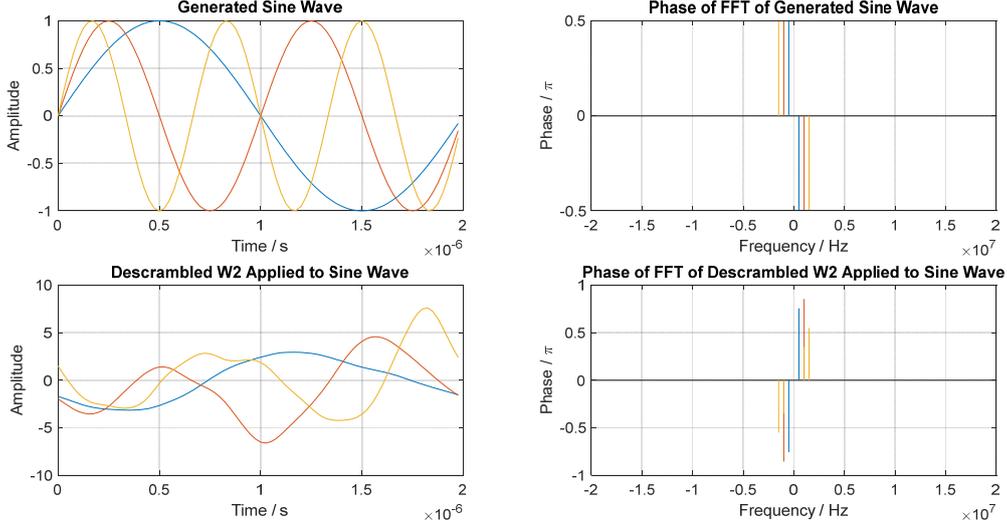

*Figure S3. Action by the descrambled form of the middle fully connected layer of DEERNet on a collection of sine waves. **Top left:** input signals (sine waves). **Top right:** input phases. **Bottom left:** output signals. Bottom right: output phases - note the inversion.*

The descrambled weight matrix appears to be inverting the phases of the harmonics that it receives (Figure S3). However, this inversion is frequency-dependent: within the pass band of the digital filter in the previous fully connected layer, the matrix in Figure S4 is antidiagonal for some frequencies, but diagonal for others. Initially, this is puzzling – up to some tenacious noise that survived the training, this corresponds to trivial phase rotations and inversions.

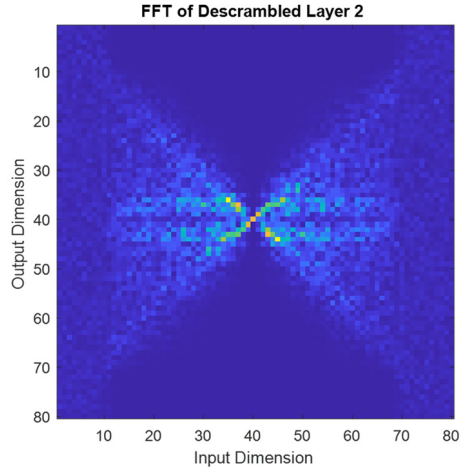

*Figure S4. Diagonal-antidiagonal structure of the Fourier transform of the descrambled middle layer weight matrix in the three-layer DEERNet.*

The essential need for phase inversions becomes clear once the group-theoretical nature of the DEER signal processing problem is considered. The integral transform in Eq (9) of the main text is, in some Euclidean metric, orthogonal. For input vectors of dimension $n$, the corresponding group is $O(n)$, and we are asking the network to approximate a particular element of this group. However, $O(n)$ is not a path-connected manifold because it has a discrete factor:

$$O(n) = SO(n) \otimes I \tag{S.16}$$

where $SO(n)$ is the special orthogonal group (which is path-connected), and $I$ is the inversion group. Loosely speaking, there are two disconnected "copies" of $SO(n)$ in $O(n)$, and the transformation that the network needs to approximate would belong to one of these copies.

Consider the situation when the solution belongs to one instance of $SO(n)$, but the initial guess randomly drops the network into an approximation of the other instance. Because $O(n)$ is not path-connected, there is no way that a continuous function optimiser like gradient descent would be able jump from one copy to the other – unless the middle layer evolves an inversion component, which appears here to come embedded into a representation of $U(1)$ group of phase rotations. The network is apparently making use of the fact $SO(n) \otimes U(1)$ is path-connected, and that:

$$O(n) \subset SO(n) \otimes U(1) \tag{S.17}$$

Thus the role of the middle fully connected layer in this larger DEERNet appears to be topological and group theoretical – the layer is a bridge between two disconnected components of the orthogonal group. The test of correctness of this hypothesis would be the sign of the determinant of the weight matrix – a direct inspection confirms that the determinant is negative.

## S6. Descrambling the input layer of an acoustic filter network

A fully connected neural network designed to remove additive noise from recordings of human voice (*5*) is included as an example with the Deep Learning Toolbox of Matlab 2020a (*3*). Its training set comes from the Common Voice database (*6*). With the exception of bias vectors (which we removed from the network architecture before training to facilitate subsequent descrambling), the network was trained as described in Matlab R2020a documentation (*3*).

The input a [129 frequency points]×[8 time points] block of the short-time Fourier transform of the input signal, stretched column-wise into a 1032-element vector. Because the network receives frequency domain data, its input and intermediate signals are not expected to be smooth – the descrambling target functional must be different from the one used for DEERNet.

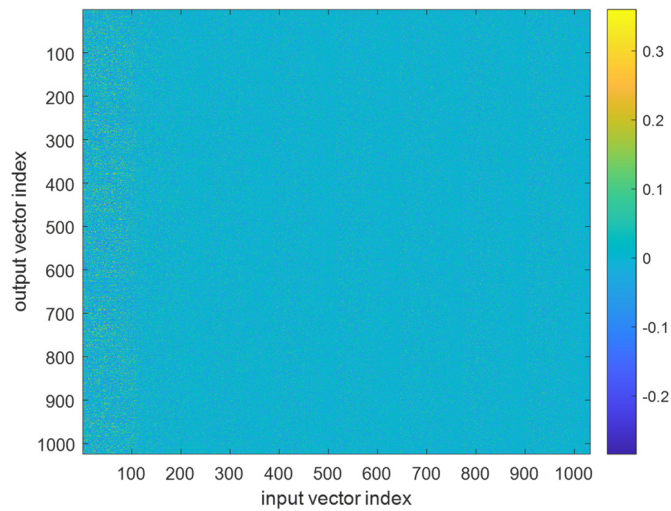

*Figure S5.* First fully connected layer weight matrix of our acoustic filter network, showing the characteristic absence of directly visible.

The immediate appearance of the weight matrix has no identifiable characteristics (Figure S5). This is typical for fully connected layers. However, the input dimension is expected to have a block structure matching the eight blocks of the input vector. This is visible in the row element autocorrelation function (Figure S6), which suggests a faint repeating pattern of 129 elements.

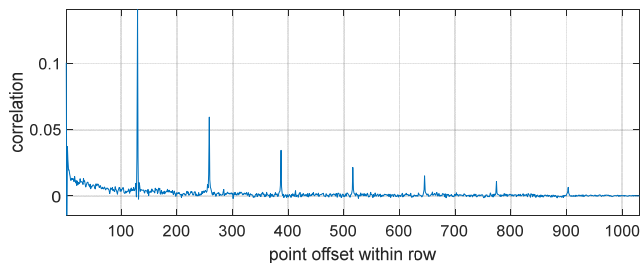

*Figure S6.* The average (over all rows) autocorrelation function of the row elements of the weight matrix of the first fully connected layer of the acoustic filter network.

The inspection of the weight matrix average over the eight correlated blocks reveals an elaborate frequency filter (Figure S7): some input frequencies are mostly rejected, and linear combinations of others are sent to the output. This suggests that the function of the matrix is to attenuate undesired frequencies, meaning that a maximum diagonality descrambler (Section S2) would be appropriate here.

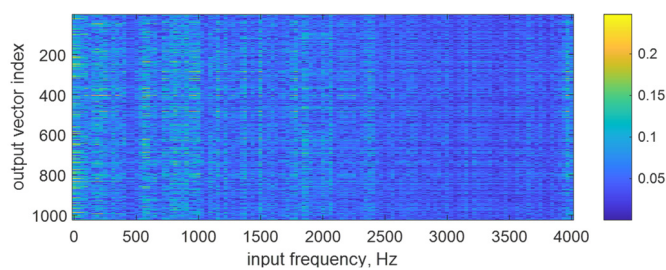

*Figure S7.* Input vector block average of the DenoiseNet input layer weight matrix with insignificant singular values filtered out.

Descrambling using MDNS criterion reveals that selective frequency attenuation is not the only thing this layer does – only about 50% of the Frobenius norm of the weight matrix in Figure S5 is condensed to the diagonal by an orthogonal transformation acting on the output side (Figure S8, left panel). The rest of the norm accumulates on the diagonals of other STFT blocks (Figure S8, right panel).

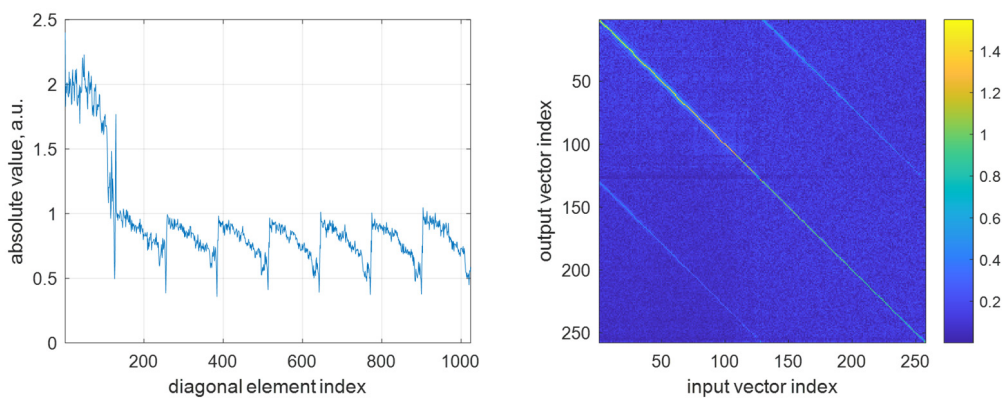

*Figure S8.* Left panel: diagonal elements of the DenoiseNet input layer after descrambling using maximum diagonality criterion. Right panel: a zoom into the top left quarter of the descrambled weight matrix.

Thus, the descrambled weight matrix is multi-diagonal: the left panel of Figure S8 reveals that the main diagonal is applying a soft low pass filter to every STFT block, and the presence of other diagonals appears – at least algebraically – to correspond to a weighed averaging and/or cancellation of individual frequencies between the blocks. The exact nature of what is happening here will be of interest to DSP specialists, this paper is not the place to investigate further. The salient point is that descrambling has again revealed interpretable structure in what previously (Figure S5) looked like noise.